\documentclass[letterpaper, articletitle=true]{article}

\usepackage[T1]{fontenc}

\usepackage{geometry}
\usepackage{setspace}
\usepackage{multicol}
\usepackage{achemso}

\usepackage{graphicx}
\usepackage{float}
\newfloat{scheme}{htbp}{los}
\floatname{scheme}{Scheme}
\floatname{chart}{Chart}
\newfloat{graph}{htbp}{loh}

\usepackage{chemformula} 
\usepackage[version = 4]{mhchem} 

\usepackage{authblk}
\author[1]{Nima Shakourifar}
\affil[1]{Department of Civil Engineering, McMaster University, Hamilton, ON, Canada}
\author[2]{Nana Ofori-Opoku}
\affil[2]{Department of Materials Science and Engineering, McMaster University, Hamilton, ON, Canada}
\author[1]{Benzhong Zhao*}

\title{Implicit Nucleation and Competitive Dynamics of \\Electrogenerated Hydrogen Nanobubbles}
\date{*Email: robinzhao@mcmaster.ca}

\begin{document}

\maketitle

\begin{abstract}
Electrogenerated gas nanobubbles strongly influence the performance of electrochemical energy-conversion systems, yet their nucleation and early evolution remain poorly understood due to fundamental limitations of existing experimental and computational approaches. Operando imaging lacks the temporal resolution required to capture nucleation events, while molecular dynamics simulations are restricted to nanometer-scale domains containing at most a few bubbles. Here, we develop a thermodynamically consistent phase-field framework that unifies dissolved-gas transport, curvature-dependent interfacial thermodynamics, and implicit bubble nucleation within a single continuum description. Using hydrogen nanobubble formation during electrocatalysis as a canonical test case, the model captures nucleation without prescribing nuclei, resolves diffusion-controlled growth under strong curvature effects, and remains computationally tractable at the micrometer scale despite hydrogen’s extremely low solubility in water. Simulations reveal how nanobubble nucleation occurs once a local supersaturation threshold is exceeded, triggering a reorganization of the chemical-potential field that focuses dissolved gas toward the nascent bubble. In multi-catalyst systems, overlapping diffusion fields lead to strong bubble–bubble interactions, including competitive growth, Ostwald ripening, and source occlusion. Extending the framework to dispersed catalyst populations shows that nanobubble survival is governed not only by catalyst size but also by spatial arrangement and diffusive competition, such that only a subset of bubbles persist while others dissolve and act as transient feeders. These results reframe electrogenerated nanobubbles as emergent, spatially organized features rather than unavoidable parasitic byproducts, and point toward electrode designs that deliberately control where bubbles nucleate and grow to preserve active area and mitigate transport losses.
\end{abstract}

\vspace{1\baselineskip}
\begin{multicols}{2}
	

\noindent Surface-adhered gas nanobubbles are a defining microscopic feature of gas-evolving electrocatalysis and a key regulator of catalytic activity. They nucleate at solid-liquid interfaces under local supersaturation and represent the earliest precursors to the micro- and macro-scale bubbles that ultimately govern electrochemical performance.\cite{anantharaj2016recent,perez2019mechanisms,roger2017earth,voiry2018low,zhao2019gas} By introducing a third, insulating phase between the electrode and electrolyte, nanobubbles deactivate active sites, disrupt ionic conduction pathways, and contribute directly to current losses and increased ohmic resistance.\cite{lu2025unveiling,qiu2024impact,araujo2024alkaline} Although reducing bubble coverage consistently improves device efficiency,\cite{foroughi2022understanding,deng2023quantitative,ross2025impact} the earliest nanoscale events---nanobubble nucleation, interaction, and selective growth---remain poorly resolved.

These efficiency penalties arise from the coupled spatiotemporal evolution of nanobubbles at catalytic sites. Limited gas solubility and diffusivity drive local supersaturation and the nucleation of surface-bound nanobubbles, which then grow, merge, or dissolve depending on curvature-controlled Laplace pressures and local concentration fields. Even during their short residence, submicron bubbles hinder electrolyte access, elevate overpotentials, and reshape mass-transport pathways, increasing the energetic cost of operation.\cite{lu2025unveiling,das2024modulating,kalita2024spontaneous,park2024insights} Resolving these earliest stages is therefore essential for predicting bubble coverage, preserving electrochemically active area, and developing strategies to mitigate performance losses.

Experimentally capturing nanobubble nucleation and early growth remains extremely challenging because it requires nanometer spatial resolution and nanosecond-microsecond temporal resolution.\cite{perez2019mechanisms} Although recent \emph{operando} imaging techniques achieve the spatial resolution necessary to track late-stage growth across fields containing tens to hundreds of nanobubbles---typically at millisecond frame rates,\cite{lemineur2021imaging,wang2020bubble}---their temporal resolution remains too coarse to measure nucleation kinetics at individual catalytic sites. As a result, most \emph{operando} studies probe nanobubble behavior only after nucleation has already occurred.

A particularly noteworthy example is the work of Lemineur \textit{et~al.},\cite{lemineur2021imaging} who employed interference reflection microscopy to monitor \emph{in operando} H$_2$ nanobubble formation on ensembles of individual Pt nanoparticles under electrochemical activation. Their measurements revealed rapid electrical isolation of nanoparticles as nanobubbles emerged and grew. Even more intriguingly, only a subset of nanoparticles produced bubbles---a behavior attributed to variations in electrochemical activity among particles and to bubble–bubble ``cross-talk'' governed by overlapping depletion fields and competitive growth. However, the method lacked the temporal resolution needed to resolve the onset of nucleation itself.

Molecular dynamics (MD) simulations, by contrast, access the requisite nanoscopic time scales to resolve nucleation, but are limited to simulation domains only a few nanometers in size. As a result, MD can capture the birth of isolated nuclei but cannot probe early-time nanobubble-nanobubble interactions or the curvature-driven competition that emerges in environments containing multiple nascent bubbles. Taken together, these limitations leave a critical spatiotemporal gap in our understanding of how nanobubbles nucleate, compete, and select during the earliest stages of gas evolution in electrochemical systems. 

To address this gap, we develop a thermodynamically consistent phase-field framework for electrocatalytic bubble nucleation and early growth. In this continuum formulation, the gas-liquid interface is represented as a finite-thickness transition layer defined by a free-energy functional, so the interface is not explicitly tracked but instead emerges naturally from the governing evolution equations.\cite{gomez2019review} Because steep gradients are confined to this narrow interfacial region while bulk fields vary smoothly, high spatial resolution is required only locally near bubble interfaces. As a result, the method remains computationally tractable on micrometer-scale domains while still resolving nanobubble-scale interfacial physics. By enabling implicit nucleation when local supersaturation crosses the free-energy barrier,\cite{ode2021thermal} our model provides the first thermodynamically consistent continuum description that unifies pre-nucleation gas buildup, nucleation kinetics, and early-to-late bubble growth under sustained electrogeneration. This capability closes a longstanding spatiotemporal gap between \emph{operando} imaging (which cannot resolve nucleation), and molecular simulations (which cannot access multi-bubble domains), offering a new route for probing nanobubble selection and competition at catalytic interfaces.

\section{Results and discussion}

Without loss of generality, we focus on the electrochemical generation of hydrogen nanobubbles as a canonical test case for our framework, motivated by both its practical importance for renewable hydrogen production and the intrinsic modeling challenges posed by hydrogen’s exceptionally low solubility in water.\cite{voiry2018low,fernandez2003henry,plyasunov2000infinite} For clarity and computational tractability, we consider an idealized Hele–Shaw geometry, which captures the essential physics of gas evolution in confined catalytic environments.\cite{saffman1958penetration} Catalyst particles are modeled as finite-area sources that impose a constant areal flux of dissolved hydrogen wherever they remain exposed to the liquid electrolyte; any portion occluded by gas ceases production, reflecting electrochemical deactivation under bubble coverage. \cite{luo2013electrogeneration,perez2019mechanisms} 

Our phase-field formulation advances a conserved Cahn–Hilliard equation for the hydrogen mole fraction $c$ and its variational chemical potential $\psi$ (energy density, $\mathrm{J}/\mathrm{m}^{3}$), coupled to a nonconserved Allen–Cahn equation for the gas volume fraction $\phi$. At local equilibrium, $c$ is essentially uniform within each bulk phase---corresponding to $\phi\approx0$ in the liquid and $\phi\approx1$ in the gas---and varies smoothly only across the diffuse interfacial region. A single free-energy functional provides the thermodynamic backbone of the model, integrating bulk and interfacial energetics, partial miscibility between hydrogen and water, and implicit heterogeneous nucleation triggered when local supersaturation exceeds a critical threshold. The extremely low solubility of hydrogen in water necessitates a free-energy landscape with strongly asymmetric equilibrium compositions and large, stable chemical-potential contrasts, while avoiding unphysical interfacial mixing.\cite{abels2012thermodynamically,ding2007diffuse} Constructing such a landscape is nontrivial, since the thermodynamic characterization of H$_2$(aq) is largely limited to infinite-dilution solubility data. Key quantities needed for a full free-energy description---such as mixing enthalpies, activity coefficients, and other excess thermodynamic properties---are not well constrained and are often inferred indirectly from temperature-dependent solubility correlations.\cite{fernandez2003henry,plyasunov2000infinite} Full mathematical details are provided in the \emph{Methods} and \emph{Supporting Information}.

\begin{figure*}[t!]
  \centering
  \includegraphics[width=0.9\textwidth]{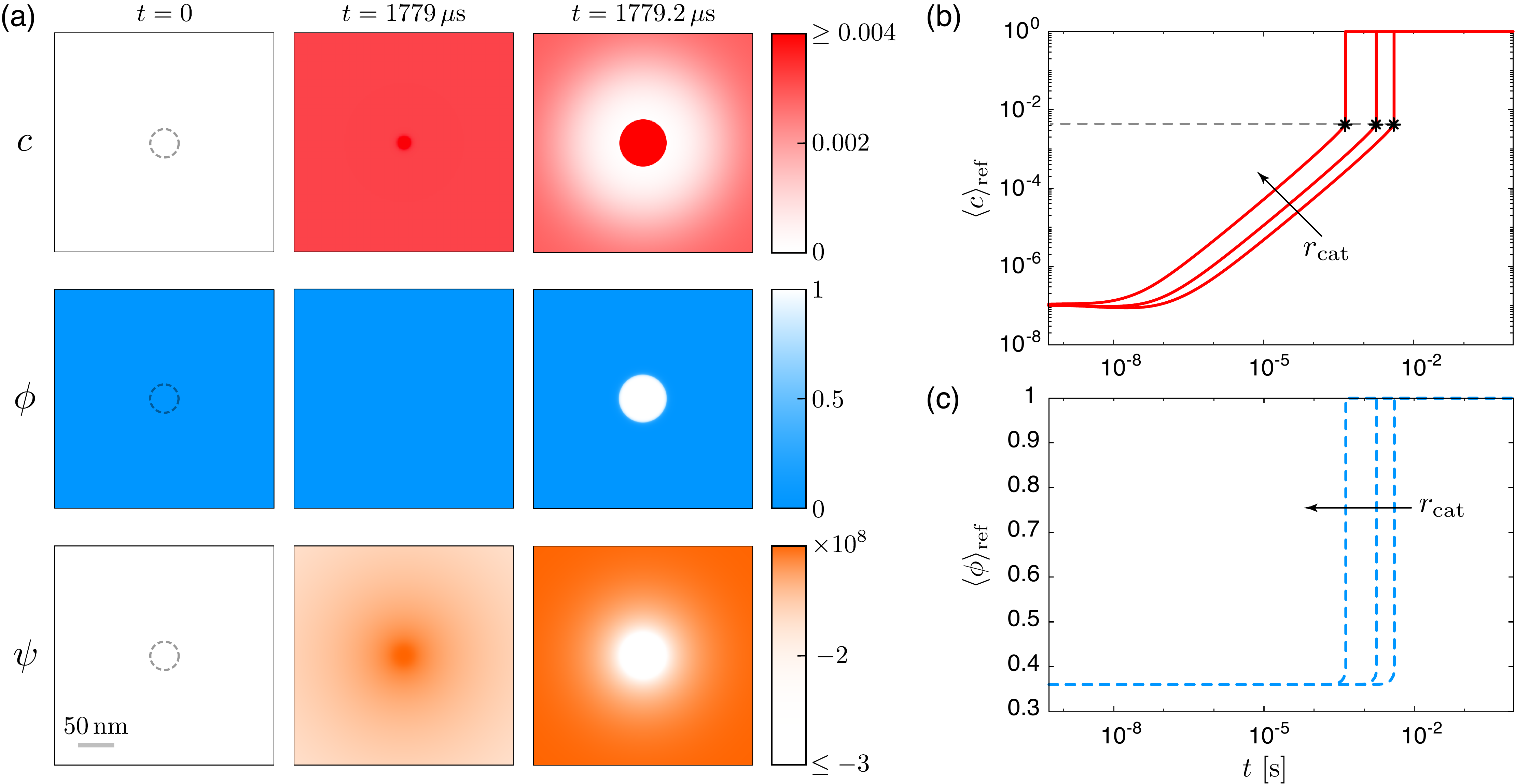}
  \caption{Implicit nucleation of a hydrogen nanobubble on a single hydrogen-producing catalyst. (a) Snapshots of the hydrogen mole fraction 
$c$ (top), gas volume fraction $\phi$ (middle), and variational chemical potential $\psi$ (bottom) at initial, pre-nucleation, and post-nucleation times. The dashed circle denotes the 15 nm catalyst footprint. After nucleation, $\psi$ reorganizes and its gradient reverses relative to the pre-nucleation state, drawing dissolved hydrogen toward the newly formed gas-liquid interface. Consequently, the $c$ field develops a distinct halo surrounding the bubble, delineating the diffusion zone that feeds its growth. (b) Evolution of the area-averaged hydrogen mole fraction $\langle c \rangle_\text{ref}$ for catalyst radii $r_\text{cat}=10,15,30$~nm. Nucleation (asterisks) occurs at a heterogeneous supersaturation threshold of order $300\times$ the equilibrium solubility and is effectively independent of catalyst size. (c) Evolution of the area-averaged gas volume fraction $\langle \phi \rangle_\text{ref}$ for the same catalyst sizes. Larger catalysts nucleate earlier, consistent with their higher hydrogen production rates and faster local buildup of dissolved gas.}
    \label{fig:single}
\end{figure*}

To establish physical consistency, we validate our phase-field model using the classical problem of bubble growth in a uniformly supersaturated liquid, which admits the well-known analytical solution of Epstein and Plesset \cite{epstein1950stability,penas2017diffusion} (\emph{Methods}). In this setting, bubble growth is controlled by time-dependent mass transfer across a curved gas–liquid interface. The Kelvin (Gibbs–Thomson) relation sets the interfacial chemical potential and local equilibrium solubility, causing smaller, high-curvature bubbles to experience larger Laplace pressures and correspondingly higher equilibrium dissolved-gas concentrations. In the Epstein–Plesset formulation, capillarity enters by prescribing the interfacial equilibrium concentration as a function of the instantaneous bubble radius, with mass transfer governed by the resulting concentration gradient in the liquid.\cite{epstein1950stability} In contrast, our approach is chemical-potential based: diffusive transport is driven by gradients in the variational chemical potential $\psi$, so the interface need not be tracked explicitly and $R(t)$ is obtained \emph{a posteriori} from the evolving phase field. As a benchmark, we compare our chemical-potential-based phase-field framework with the Epstein-Plesset concentration-gradient theory under capillarity; our simulations show excellent agreement with the Epstein-Plesset growth law over its range of validity, confirming that the model correctly captures diffusion-controlled bubble growth under finite interfacial curvature (Figure~\ref{fig:ep}).

Having validated that our model reproduces diffusion-controlled bubble growth, we next address how a bubble first appears. Classical nucleation theory requires that a free-energy barrier---arising from interfacial cost versus bulk driving force---be overcome.\cite{brennen2014cavitation} We model this barrier-crossing process by adapting the fluctuation-driven, critical-phase-field-value concept originally developed for solids\cite{ode2021thermal} to gas-phase nucleation in an extremely low-solubility, partially miscible hydrogen-water system under electrogeneration. A small, localized bias in $\phi$ is applied over a tiny region at each catalyst footprint, representing an activation-scale fluctuation that perturbs the liquid phase without creating a separate phase.

\begin{figure*}[t!]
  \centering
  \includegraphics[width=0.9\textwidth]{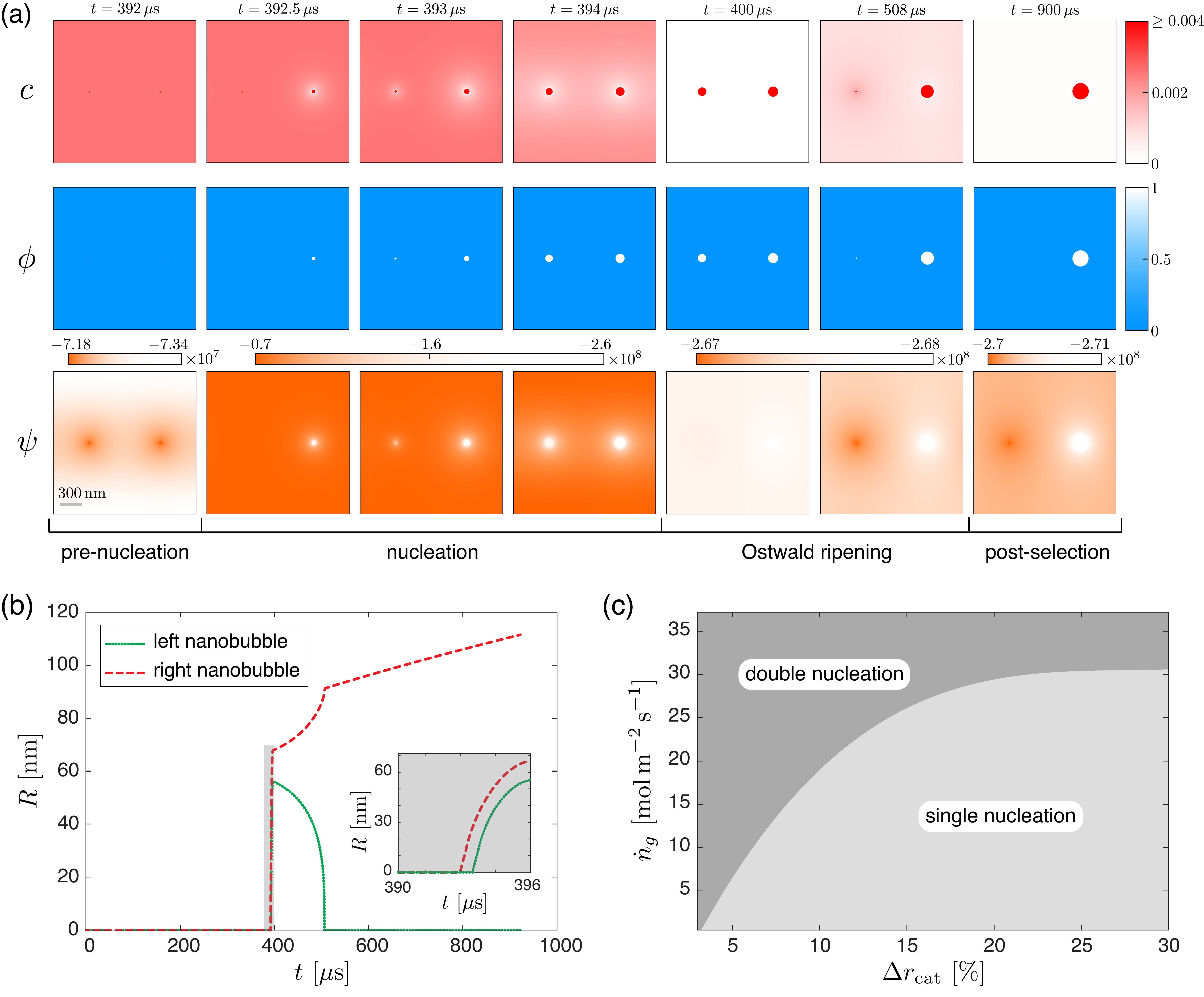}
  \caption{Nanobubble nucleation and growth on a pair of catalysts. (a) Snapshots of the hydrogen mole fraction 
$c$ (top), gas volume fraction $\phi$ (middle), and variational chemical potential $\psi$ (bottom). The dashed circles denotes the 15 nm (left) and 15.5 nm (right) catalyst footprints. Prior to nucleation, both catalysts generate dissolved H$_2$, which diffuses into the surrounding domain. Nucleation occurs first on the larger catalyst, followed by the smaller one. As the bubbles grow and fully occlude their respective footprints, hydrogen production ceases at both sites. Ostwald ripening then governs the evolution: dissolved hydrgen flows from the smaller, higher-curvature bubble to the larger one until the smaller bubble completely dissoves. Once this occurs, the left catalyst becomes re-exposed to the electrolyte and resumes hydrogen production, which is subsequently captured by the larger bubble, sustaining its continued growth. (b) Bubble radius $R$ provides quantitative signatures of this process, with insets highlighting differences between the two catalysts sites at the moment of nucleation. (c) Phase diagram delineating single- and double-nucleation regimes as a function of percentage size difference $\Delta{r_\text{cat}}$ and imposed areal hydrogen production rate $\dot{n}_g$.}
    \label{fig:double}
\end{figure*}

Nucleation occurs once the local dissolved-hydrogen concentration exceeds a supersaturation level at which the bulk free-energy gain outweighs the interfacial penalty. Below this nucleation threshold, the perturbation remains neutrally stable; above it, the bias naturally switches off as the region becomes gas-covered and the fluctuation relaxes into a stable nucleus that subsequently grows by diffusion. Guided by single-site \emph{operando} experiments and molecular dynamics simulations, we adopt a heterogeneous nucleation threshold of order $300\times$ the equilibrium solubility, consistent with measurements showing little dependence on nanoelectrode size or composition.\cite{chen2014electrochemical,german2018critical,perez2019mechanisms,chen2018hydrogen}\\ 

\noindent\textbf{Single Catalyst.} We demonstrate implicit nucleation of a hydrogen nanobubble on a single hydrogen-producing catalyst (Figure~\ref{fig:single}). An areal hydrogen production rate of $\dot{n}_g=0.5~\mathrm{mol\,m^{-2}\,s^{-1}}$ is imposed on the catalyst, representative of the lower end of experimentally reported values for isolated nanoelectrodes.\cite{luo2013electrogeneration} As hydrogen is generated, the dissolved-hydrogen mole fraction $c$ and the corresponding variational chemical potential $\psi$ rise throughout the surrounding liquid, while the gas volume fraction $\phi$ remains unchanged. Nucleation occurs at $t=1779~\mu$s, reflected by a sudden increase in $\phi$ and an abrupt reversal in the gradient of $\psi$, which directs dissolved hydrogen toward the newly formed gas-liquid interface. This reorganization of $\psi$ gives rise to a pronounced halo in the $c$ field surrounding the nascent bubble, delineating the diffusion zone that subsequently supplies its growth (Figure~\ref{fig:single}a). 

Repeating the simulation for catalyst radii of 10, 15, and 30~nm shows that nucleation consistently occurs at a supersaturation of $\sim300\times$ the equilibrium solubility, independent of catalyst size. Larger catalysts nucleate earlier in absolute time, however, owing to their higher hydrogen production rates and correspondingly faster local buildup of dissolved gas (Figure~\ref{fig:single}b).\\

\noindent\textbf{Catalyst Pair.} We next examine nanobubble nucleation and evolution on a pair of closely spaced catalysts, which display rich dynamics arising from interactions between the nascent bubbles. The two catalyst particles have slightly different radii (15 and 15.5 nm) and are positioned 1~$\mu$m apart in an initially undersaturated domain (Figure~\ref{fig:double}a). This small size difference introduces a mild asymmetry in the early concentration field, as the larger catalyst accumulates dissolved hydrogen slightly faster and therefore nucleates first (Figure~\ref{fig:double}a,b). Once nucleation occurs, the resulting chemical-potential gradient drives dissolved hydrogen toward the newly formed interface, and the nascent bubble acts as a diffusive sink. Before the associated depletion front reaches the neighboring catalyst, however, the dissolved-hydrogen concentration at the smaller catalyst also reaches the nucleation threshold, leading to a second nucleus. At this stage, both bubbles draw flux from the surrounding domain. 

As the bubbles grow and progressively occlude their respective footprints, hydrogen production on the catalyst surfaces ceases. Growth nevertheless continues because the liquid remains supersaturated, and the bubbles expand by diffusive uptake until the dissolved-hydrogen concentration relaxes toward its near-equilibrium value. Beyond this point, curvature differences dominate the local thermodynamics and trigger Ostwald ripening: the smaller bubble, having higher interfacial curvature, possesses a higher chemical potential, and material transfer proceeds from the smaller bubble to the larger one\cite{shin2015growth}. As the smaller bubble shrinks, its underlying catalyst surface becomes re-exposed to the liquid electrolyte and resumes hydrogen production, but the nearby larger bubble captures this flux, suppressing renewed nucleation at the smaller site and fueling its own growth---even though the larger catalyst still remains inactive underneath its bubble. The full evolution of the bubble pair therefore unfolds in four characteristic stages: pre-nucleation, nucleation, Ostwald ripening, and post-selection (Figure~\ref{fig:double}a).

The nucleation and growth dynamics are governed by the size contrast between the catalysts and by the imposed areal production rates. Increasing the size mismatch can delay---or even prevent---nucleation at the smaller footprint if the depletion field from the earlier-nucleating site arrives quickly enough. At higher areal production rates, however, both sites can still nucleate despite the competitive interaction. Figure~\ref{fig:double}c summarizes this behavior in a phase diagram delineating single- and double-nucleation regimes as a function of size contrast and production rate.\\

\begin{figure*}[t!]
  \centering
  \includegraphics[width=\textwidth]{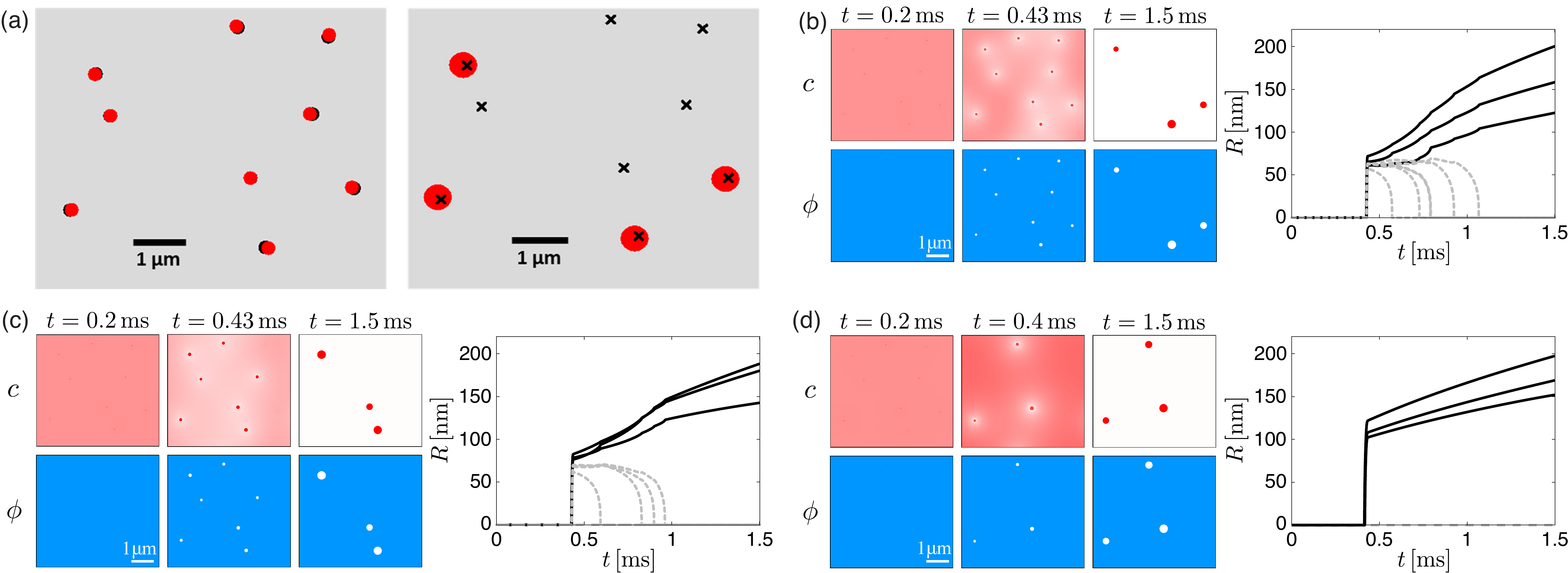}
  \caption{Nanobubble nucleation and growth in a dispersed catalyst population. (a) Interference reflection microscopy experiments on electrochemically activated Pt nanoparticles reported by Lemineur \textit{et~al.}\cite{lemineur2021imaging}. Left: nanoparticle positions before (black circles) and after (red circles) electrochemical activation. Right: nanoparticle positions (crosses) and hydrogen nanobubble locations (red circles) following electrochemical activation. Motivated by these observations, we investigate the role of catalyst size polydispersity on nanobubble nucleation and persistence by fixing the mean catalyst footprint radius at 15~nm and considering three levels of size dispersion, corresponding to standard deviations of (b) 0\thinspace\%, (c) 1.3\thinspace\%, and (d) 13\thinspace\%. For each case, the left panels show the spatiotemporal evolution of the hydrogen mole fraction $c$ and gas volume fraction $\phi$, while the right panels show the corresponding bubble radius evolution. Persistent nanobubbles are indicated by solid lines, whereas transient nanobubbles are denoted by dashed lines.}
    \label{fig:multi}
\end{figure*}

\noindent\textbf{Dispersed Catalyst Population.} Finally, we examine nanobubble nucleation and evolution in a dispersed population of catalyst particles, analogous to the interference reflection microscopy experiments on electrochemically activated Pt nanoparticles reported by Lemineur \textit{et~al.}\cite{lemineur2021imaging}. Those experiments showed that only a subset of catalyst particles ($\sim$30\thinspace\% on average) produced stable nanobubbles (Figure~\ref{fig:multi}a), a behavior hypothesized to arise from heterogeneity in electrochemical activity as well as to bubble–bubble ``cross-talk'' mediated by overlapping depletion fields and competitive growth. However, the experimental temporal resolution was insufficient to directly resolve the onset of nucleation. An additional intriguing observation was that some nanoparticles without stable nanobubbles exhibited slight lateral motion following electrochemical activation (Figure~\ref{fig:multi}a). This behavior may reflect transient, asymmetric nanobubble growth at the particle surface, followed by rapid dissolution before the bubbles could be detected.

Here, we simulate the nucleation and growth of H$_2$ nanobubbles in a $5\times 5~\mu\text{m}^2$ domain containing nine sparsely distributed catalysts, each producing dissolved hydrogen at the same areal rate of $0.5~\mathrm{mol\,m^{-2}\,s^{-1}}$. Periodic boundary conditions are imposed, so the computational domain represents a repeating tile of a much larger catalyst dispersion, consistent with the experimental field of view of Lemineur \textit{et~al.}\cite{lemineur2021imaging}. Motivated by the polydisperse nature of catalyst particles observed experimentally, we set the mean catalyst footprint radius to 15~nm and consider three levels of size dispersion, corresponding to standard deviations of 0, 1.3\thinspace\%, and 13\thinspace\%.

In simulations with perfectly monodisperse catalyst particles, all catalysts nucleate nanobubbles, albeit with slightly different onset times (Figure~\ref{fig:multi}b). Sites experiencing weaker local depletion from overlapping diffusion fields nucleate earlier and develop larger initial bubble radii than neighboring catalysts. These small differences in initial bubble size subsequently drive Ostwald ripening, leading to the dissolution of six nanobubbles over time, while three persist. As nanobubbles dissolve, the underlying catalyst sites become re-exposed and resume hydrogen production, which feeds the surviving bubbles. This renewed source-driven flux ultimately dominates over curvature-driven Ostwald ripening and facilitates the persistence of the remaining bubbles. Notably, this behavior highlights the emergent complexity of the system: the earliest nucleating bubbles are not necessarily the ones that ultimately survive, because long-term survival is controlled by sustained access to hydrogen sources and diffusive competition rather than by nucleation order alone.

Introducing modest catalyst size dispersion further amplifies this effect. For a size distribution with a standard deviation of 1.3\thinspace\%, only seven catalysts nucleate nanobubbles, as the initially formed bubbles act as strong local sinks for dissolved hydrogen, depleting the surrounding concentration field and suppressing nucleation at the remaining sites. Subsequent Ostwald ripening dissolves several of the smaller bubbles, leaving three stable survivors (Figure~\ref{fig:multi}c). The identities of the surviving bubbles differ from those in the monodisperse case, underscoring the sensitivity of bubble selection to subtle variations in catalyst size and local diffusive interactions. Further increasing the size dispersion by an order of magnitude confines nucleation to only the three largest catalysts, whose bubbles scavenge dissolved hydrogen from neighboring sites and thereby persist without dissolving (Figure~\ref{fig:multi}d).

We emphasize that the present framework is gap-averaged and quasi-two-dimensional and therefore does not explicitly resolve the three-phase contact line or associated wettability effects of surface nanobubbles. Nanobubbles are known to exhibit contact-line pinning, which leads to variations in the apparent contact angle during growth and shrinkage.\cite{weijs2013surface,lohse2015surface} In contrast, our model assumes an effective, uniform contact angle of $90^\circ$. This assumption is appropriate for the regime targeted here: the nanosecond-to-microsecond, diffusion-controlled window immediately following nucleation. During this early stage, bubble evolution is governed primarily by curvature-induced Laplace pressures, local supersaturation, and dissolved-gas transport, which together determine the nucleation threshold and initial growth kinetics. Contact-line pinning predominantly influences later-time shape evolution, residence, and detachment, and therefore does not control the nucleation barrier emphasized in this work.

\section{Conclusions}

We developed a thermodynamically consistent phase-field framework to describe the nucleation and evolution of electrogenerated gas nanobubbles, bridging a critical gap between experiments, whose temporal resolution is insufficient to resolve nucleation, and molecular dynamics simulations, which are restricted to single-bubble, nanometer-scale domains. Using electrochemical hydrogen nanobubble formation as a canonical test case, we demonstrate that the framework captures dissolved gas transport, curvature-driven interfacial thermodynamics, and implicit nucleation in systems with extremely low equilibrium gas solubility---a longstanding challenge in modeling gas–liquid electrochemical systems.

For a single catalyst, nanobubble nucleation occurs once a critical supersaturation is reached, followed by a rapid reorganization of the chemical-potential field that drives dissolved gas toward the nascent bubble (Figure~\ref{fig:single}). For a pair of catalysts with a slight size mismatch, the model resolves the complete sequence of coupled processes: pre-nucleation gas accumulation, nucleation at both sites, dissolution of the smaller bubble via Ostwald ripening, and sustained growth of the larger bubble fueled by gas produced at the neighboring catalyst (Figure~\ref{fig:double}). Extending the framework to dispersed catalyst populations reveals that both catalyst size polydispersity and spatial arrangement exert strong control over nanobubble nucleation, competition, and long-term persistence. Although larger catalysts tend to nucleate bubbles earlier, survival is governed by diffusive competition with neighboring sites. As a result, even in perfectly monodisperse systems, only a subset of nanobubbles persists, while others dissolve and act as transient feeders that supply gas to the surviving bubbles (Figure~\ref{fig:multi}).

The absence of stable bubble formation on some catalysts should therefore not be interpreted as electrochemical inactivity. On the contrary, it is often a desirable outcome: bubble-free catalysts continue to operate without losses in active area or increases in ohmic resistance associated with bubble coverage. Such behavior is expected to be common in practical electrodes, where intrinsic nanoparticle polydispersity and conventional deposition methods (e.g., drop casting and spray coating) introduce spatial heterogeneity and clustering through coffee-ring effects, naturally biasing bubble nucleation toward only a subset of sites.

These findings highlight new opportunities for rational electrode design. Catalyst distributions could be deliberately engineered to create designated ``bubble-nucleation hubs'' surrounded by arrays of feeder catalysts that remain bubble-free, combined with tailored porous transport layers that promote efficient bubble removal at the hub sites. Such architectures are experimentally accessible using precision coating and patterning techniques, including electrophoretic deposition\cite{besra2007review,corni2008electrophoretic} and pulse electrodeposition\cite{popov2001pulse,schlesinger2011modern}. More broadly, our results reframe electrogenerated bubbles not as unavoidable parasitic byproducts, but as spatially controllable features that can be deliberately exploited to enhance electrochemical performance.

\section{Methods}

\subsection{Phase-Field Modeling Framework}

We develop a thermodynamically consistent phase-field framework to describe electrogenerated gas–liquid systems in a gap-averaged Hele–Shaw geometry. The model resolves the coupled evolution of the dissolved-gas mole fraction $c$, the variational chemical potential $\psi$, and the gas volume fraction $\phi$ using a Cahn–Hilliard equation for conserved mass transport and an Allen–Cahn equation for nonconserved phase evolution. The formulation is derived from a free-energy functional $F$ that accounts for bulk thermodynamics, interfacial energetics, partial miscibility, and the nucleation barrier.

The system is treated as quasi-incompressible, with constant molar densities in each phase. Under this assumption, the governing equations read
\begin{equation}
\nu(c)\,\frac{\partial c}{\partial t}
\;-\;
\nabla\!\cdot\!\left(
M^{c}\,\nabla \frac{\delta F}{\delta c}
\right)
\;=\;0,
\end{equation}
\begin{equation}
\nu(c)\,\frac{\partial \phi}{\partial t}
\;+\;
M^{\phi}\,\frac{\delta F}{\delta \phi}
\;=\;0,
\end{equation}
where $\nu(c)$ is the mixture molar density (mol\,m$^{-3}$), and $M^{c}$ and $M^{\phi}$ are the mobilities associated with composition and phase evolution, respectively. The composition mobility is defined as
\begin{equation}
M^{c} \;=\; \frac{D}{R T}\,\bigl(0.01 + c(1-c)\bigr),
\end{equation}
where $D$ is the molecular diffusivity (m$^{2}$\,s$^{-1}$), $R$ is the ideal gas constant (J\,mol$^{-1}$\,K$^{-1}$), and $T$ is the temperature (K). The small regularization term ensures nonzero mobility in the dilute limits. The phase mobility is given by
\begin{equation}
M^{\phi} \;=\; \frac{D}{R T b^{2}},
\end{equation}
where $b$ is the gap thickness of the Hele-Shaw cell.

\begin{figure} [H]
  \centering
  \includegraphics[width=0.95\columnwidth]{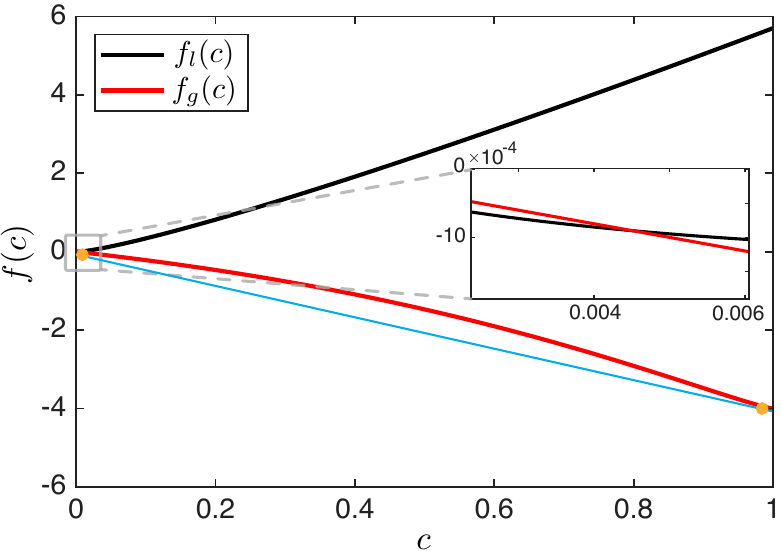}
  \caption{Bulk excess free-energy densities for the pure liquid (water) phase, $f_l(c)$ (black), and the pure gas (hydrogen) phase, $f_g(c)$ (red), as functions of the hydrogen mole fraction $c$. The common-tangent construction (blue) identifies the equilibrium coexistence compositions in the liquid and gas phases (orange circles). The inset highlights the vicinity of the intersection $f_l(c)=f_g(c)$, which marks the composition at which the global bulk free-energy minimum shifts from liquid-dominated to gas-dominated, providing an estimate of the thermodynamic crossover relevant to nucleation.}
  \label{fig:flfg}
\end{figure}

The quasi-incompressible closure follows Jacqmin,\cite{Jacqmin1999} with the mixture molar density expressed as the harmonic average of the phase molar densities,
\begin{equation}
\frac{1}{\nu(c)} \;=\; \frac{c}{\nu_{g}} + \frac{1-c}{\nu_{l}},
\end{equation}
where $\nu_{g}$ and $\nu_{l}$ denote the constant molar densities of the gas and liquid phases, respectively.

The free-energy functional combines interfacial gradient penalties, a double-well barrier in the phase field, and a composition-dependent excess mixing contribution adapted from binary-alloy phase-field formulations:\cite{CahnHilliard1959,fu2016thermodynamic}
\begin{equation}
\begin{aligned}
F \;=\; \int_{V} \bigg[
&\frac{\epsilon_{\phi}^{2}}{2}\,|\nabla \phi|^{2}
+\frac{\epsilon_{c}^{2}}{2}\,|\nabla c|^{2}
+\omega\,W(\phi) \\
&\quad
+\omega_{\mathrm{mix}}
\Big\{
f_{g}(c)\,g(\phi)
+f_{l}(c)\,[1-g(\phi)]
\Big\}
\bigg]\;\mathrm{d}V .
\end{aligned}
\label{eq:free_energy}
\end{equation}

Here, $\epsilon_{\phi}$ and $\epsilon_{c}$ control the thickness and energetic cost of the diffuse interfaces in the phase field and composition, respectively. The function $W(\phi)=\tfrac14\,\phi^{2}(1-\phi)^{2}$ is a symmetric double-well potential that enforces phase separation and introduces a nucleation barrier, scaled by $\omega$ (energy per unit volume). The bulk free-energy densities $f_{g}(c)$ and $f_{l}(c)$ describe hydrogen thermodynamics in the gas and liquid phases and are interpolated smoothly across the interface using
$g(\phi)=-\phi^{2}(2\phi-3)$, which satisfies $g(0)=0$, $g(1)=1$, and has zero slope at both stable phases.

Binary-alloy free-energy forms are well suited to this problem because extremely low-solubility gas–liquid systems are thermodynamically analogous to highly asymmetric binary mixtures, in which one component partitions almost entirely into one phase while remaining only at trace concentrations in the other.\cite{provatas2011phase} In the free-energy density inside the integral of Equation~\ref{eq:free_energy}, the first two terms penalize spatial gradients in the phase field and composition, respectively, with $\epsilon_{\phi}^{2}$ and $\epsilon_{c}^{2}$ setting the characteristic interfacial thicknesses and energetic costs.
The third term provides the energetic barrier required for heterogeneous nucleation. The fourth term represents the excess bulk mixing free energy, interpolating between liquid and gas branches through $g(\phi)$.

The \textit{ad hoc} excess free-energy functions are designed following a hybrid Wilson–Landau form\cite{wilson1964vapor,landau2013statistical} to capture the essential thermodynamic asymmetry of the hydrogen–water system and to construct a free-energy landscape suitable for threshold-based nucleation:
$f_{l}(c)=c\ln(300\,c)$ and
$f_{g}(c)=(1-c)\ln(1-c)-c(c+1)^{2}$.
A common-tangent construction identifies the equilibrium compositions in each phase (Figure~\ref{fig:flfg}), while the crossing of the two branches indicates the transition between stable states. Together, the bulk mixing free energy and the interfacial double-well term define a free-energy barrier separating the homogeneous liquid state from a stable gas nucleus. This barrier plays the role of the classical nucleation barrier and is crossed implicitly when local supersaturation is sufficient, without prescribing a critical nucleus size \emph{a priori}.

For parameterization of the interfacial terms, we set $\epsilon_{\phi}^{2}=\gamma\,b$, where $\gamma=0.072~\mathrm{N\,m^{-1}}$ is the hydrogen–water surface tension and $b$ is the Hele–Shaw gap thickness. This choice ensures that the phase-field gradient contribution reproduces the correct capillary energy when the three-dimensional free energy is gap-averaged, and that the sharp-interface surface tension is recovered in the appropriate thin-interface limit. 

The composition gradient coefficient is linked via $\epsilon_{c}^{2}=\epsilon\,\epsilon_{\phi}^{2}$, where $\epsilon$ is a dimensionless calibration parameter that controls the relative weighting of interfacial versus bulk energetic contributions and thereby tunes the miscibility contrast between the phases. For prescribed diffuse-interface thicknesses $\sigma_{\phi}$ and $\sigma_{c}$, the energetic barriers associated with phase separation and composition partitioning are set by
$\omega=\epsilon_{\phi}^{2}/\sigma_{\phi}^{2}$ and
$\omega_{\mathrm{mix}}=\epsilon_{c}^{2}/\sigma_{c}^{2}$,
respectively. Together, these parameters determine the height of the free-energy barrier separating metastable liquid states from stable gas nuclei, and therefore control the nucleation threshold in the model.

Transport is governed by the diffusivity of hydrogen in water at ambient conditions, taken as $D=5\times10^{-9}~\mathrm{m^{2}\,s^{-1}}$.\cite{ferrell1967diffusion}
For thermodynamic scaling, the molar densities are fixed at $500~\mathrm{mol\,m^{-3}}$ for hydrogen---referenced to the mean pressure inside bubbles throughout this work---and $56{,}000~\mathrm{mol\,m^{-3}}$ for liquid water.

With the free-energy functional $F$ defined, the variational chemical potential driving dissolved-gas transport is obtained as
\begin{equation}
\psi \;\equiv\; \frac{\delta F}{\delta c}
\;=\; \omega_{\mathrm{mix}}\Big[ f'_{l}(c)\,[1-g(\phi)] + f'_{g}(c)\,g(\phi) \Big]
\;-\; \epsilon_{c}^{2}\,\nabla^{2}c,
\end{equation}
which combines bulk thermodynamic driving forces with a gradient penalty that regularizes sharp concentration variations across the diffuse gas–liquid interface.

\subsection{Model Validation}

To validate the phase-field framework, we consider the classical benchmark problem of diffusion-controlled growth of a gas bubble in a quiescent, uniformly supersaturated liquid. The bubble is placed at the center of a sufficiently large domain with an imposed supersaturation of approximately $300\times$ the equilibrium dissolved-hydrogen concentration ($S\approx300$). At ambient conditions, the equilibrium solubility of hydrogen in water corresponds to a mole fraction $c_{\mathrm{eq}}\approx1.4\times10^{-5}$,\cite{wiebe1932solubility} so the imposed supersaturation corresponds to an initial dissolved-hydrogen mole fraction $c\approx4.2\times10^{-3}$.\cite{luo2013electrogeneration}

\begin{figure}[H] 
  \centering
  \includegraphics[width=0.95\columnwidth]{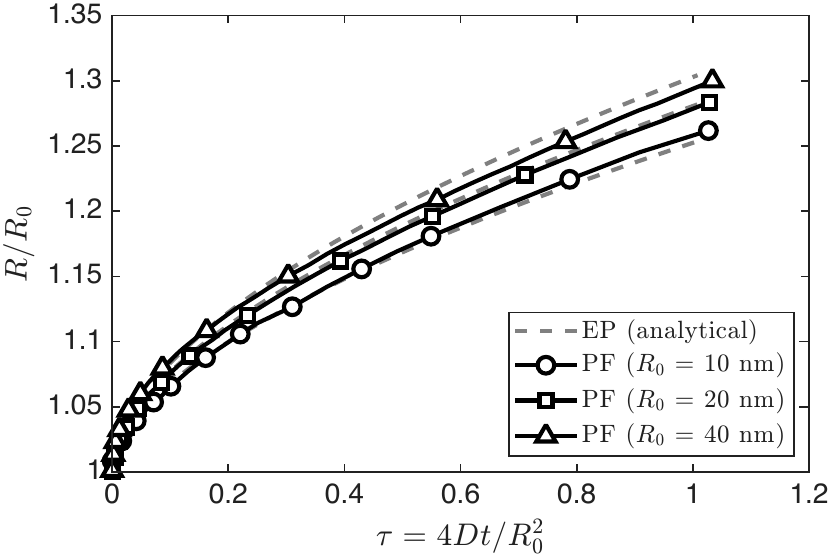}
  \caption{Diffusion-controlled growth of H$_2$ nanobubbles with initial radii of 10, 20, and 40~nm in a large, uniformly supersaturated aqueous bath (\(S \approx 300\)). Phase-field predictions (solid lines with symbols) show excellent agreement with the Epstein–Plesset analytical solution (gray dashed line) within its regime of validity, corresponding to early times \(\tau = 4Dt/R_{0}^{2} \ll 1\).}
  \label{fig:ep}
\end{figure}

We compare phase-field simulations for several initial bubble radii $R_{0}$ with the Epstein-Plesset analytical solution. Because the model is formulated in a gap-averaged, quasi-two-dimensional Hele–Shaw geometry, the simulated bubbles are effectively cylindrical rather than spherical; we therefore employ the corresponding cylindrical form of the Epstein–Plesset solution (see \emph{Supporting Information}). Varying $R_{0}$ probes the curvature sensitivity imposed by the Kelvin relation: smaller bubbles exhibit higher interfacial chemical potentials and thus higher equilibrium dissolved-gas concentrations at the interface. The excellent agreement across radii demonstrates that the gradient-energy representation of capillarity in the phase-field model correctly reproduces curvature-controlled, diffusion-limited bubble growth (Figure~\ref{fig:ep}).

The Epstein–Plesset solution is valid in the early-time regime \(4Dt/R_{0}^{2}\ll1\),\cite{penas2017diffusion} where growth is governed by transient diffusion in the liquid. In the Epstein–Plesset formulation, capillarity enters explicitly through the radius-dependent interfacial equilibrium concentration, and mass transfer is driven by the resulting concentration gradient.\cite{epstein1950stability} By contrast, in our chemical-potential-based framework, diffusive transport is driven by gradients of the variational chemical potential $\psi$. The gas–liquid interface is not tracked explicitly, and the bubble radius $R(t)$ is obtained \emph{a posteriori} from the evolving phase field.

\section*{Acknowledgements}

The authors thank the Natural Sciences and Engineering Research Council of Canada (NSERC) for supporting this research through Alliance Grants (ALLRP-563682-21) and Discovery Grants (RGPIN-2019-07162). The authors also thank ChungHyuk Lee and Drew Higgins for helpful discussions and comments.

\section*{Supporting Information}

The following files are available free of charge.
\begin{itemize}
  \item Supporting Information (PDF): Model calibration and validation; numerical implementation details, including the implicit nanobubble nucleation procedure.
  \item Movies S1-S3: Time-resolved evolution of $\phi$ and $c$ for a single catalyst, a catalyst pair, and a dispersed catalyst population.
\end{itemize}

\bibliography{references}

\end{multicols}

%
%
%
%
%

\end{document}